\documentclass[sigconf]{acmart}

\usepackage[UKenglish]{babel}
\usepackage{cleveref}
\usepackage{subfigure}

\usepackage{multirow}

\AtBeginDocument{%
  }
\usepackage{xcolor}

\copyrightyear{2025}
\acmYear{2025}
\setcopyright{cc}
\setcctype{by}
\acmConference[E-ENERGY '25]{The 16th ACM International Conference on Future and Sustainable Energy Systems}{June 17--20, 2025}{Rotterdam, Netherlands}
\acmBooktitle{The 16th ACM International Conference on Future and Sustainable Energy Systems (E-ENERGY '25), June 17--20, 2025, Rotterdam, Netherlands}\acmDOI{10.1145/3679240.3734587}
\acmISBN{979-8-4007-1125-1/2025/06}

\begin{document}

\title{Human-in-the-Loop AI for HVAC Management Enhancing Comfort and Energy Efficiency}


\author{Xinyu Liang}
\email{xinyu.liang@monash.edu}
\orcid{0000-0001-7442-3637}
\affiliation{%
  \institution{Monash University and OPTIMA}
  \city{Melbourne}
  \state{Victoria}
  \country{Australia}
}

\author{Frits de Nijs}
\email{frits.nijs@monash.edu}
\orcid{0000-0003-4466-2447}
\affiliation{%
  \institution{Monash University}
  \city{Melbourne}
  \state{Victoria}
  \country{Australia}
}

\author{Buser Say}
\email{buser.say@monash.edu}
\orcid{0000-0003-2822-5909}
\affiliation{%
  \institution{Monash University}
  \city{Melbourne}
  \state{Victoria}
  \country{Australia}
}

\author{Hao Wang}
\authornote{Corresponding author.}
\email{hao.wang2@monash.edu}
\orcid{0000-0001-5182-7938}

\affiliation{%
  \institution{Monash University}
  \city{Melbourne}
  \state{Victoria}
  \country{Australia}
}


\renewcommand{\shortauthors}{Liang et al.}

\begin{abstract}
Heating, Ventilation, and Air Conditioning (HVAC) systems account for approximately 38\% of building energy consumption globally, making them one of the most energy-intensive services. The increasing emphasis on energy efficiency and sustainability, combined with the need for enhanced occupant comfort, presents a significant challenge for traditional HVAC systems. These systems often fail to dynamically adjust to real-time changes in electricity market rates or individual comfort preferences, leading to increased energy costs and reduced comfort. In response, we propose a Human-in-the-Loop (HITL) Artificial Intelligence framework that optimizes HVAC performance by incorporating real-time user feedback and responding to fluctuating electricity prices. Unlike conventional systems that require predefined information about occupancy or comfort levels, our approach learns and adapts based on ongoing user input. By integrating the occupancy prediction model with reinforcement learning, the system improves operational efficiency and reduces energy costs in line with electricity market dynamics, thereby contributing to demand response initiatives. Through simulations, we demonstrate that our method achieves significant cost reductions compared to baseline approaches while maintaining or enhancing occupant comfort. This feedback-driven approach ensures personalized comfort control without the need for predefined settings, offering a scalable solution that balances individual preferences with economic and environmental goals.
\end{abstract}



\maketitle

\section{Introduction}
Heating, Ventilation, and Air Conditioning (HVAC) systems are essential for providing thermal comfort in various settings, but they are also among the most significant energy consumers globally, accounting for approximately 38\% of total energy consumption in buildings across sectors~\cite{gonzalez2022review}. As global energy demands rise and sustainability initiatives become more important, optimizing HVAC systems has become a priority for reducing both operational costs and environmental impact~\cite{han2019review, yao2021state}. To address this, advanced control mechanisms for HVAC systems that integrate real-time energy and environmental dynamics, and occupant comfort are crucial. This paper explores intelligent control mechanisms of HVAC that can dynamically balance energy efficiency and user comfort.

In addition to providing comfort, energy efficiency of HVAC systems is also critical due to their high energy consumption. Traditional HVAC systems typically rely on fixed settings that do not account for varying comfort needs or changes in occupancy, leading to inefficiencies. These limitations result in higher operating costs and reduced occupant comfort~\cite{wei2018review, zhou2023comprehensive}. Given the inability of traditional HVAC systems to adapt to fluctuating energy prices, grid dynamics and diverse occupant comfort needs, Intelligent Energy Management Systems (IEMS) have emerged as a promising solution. IEMS are designed to optimize energy consumption while maintaining comfortable indoor environments by incorporating real-time data and control mechanisms. Based on the control strategy, IEMS can typically be categorized into two main approaches: direct control and indirect control.

Indirect control systems in IEMS focus on influencing user behavior through recommendations, rather than directly adjusting HVAC settings. These systems monitor environmental conditions, energy usage patterns, and localized renewable generation such as PV to provide feedback and suggestions to users, encouraging them to adopt more energy-efficient habits~\cite{alsalemi2019endorsing,popa2019deep,sardianos2020real,sardianos2020rehab,kar2019revicee}. A key advantage is fostering long-term behavioral change, helping users become more eco-friendly while being cost-effective and adaptable to various HVAC systems~\cite{alsalemi2019endorsing,popa2019deep}. However, these systems rely on user engagement to manually adjust settings, which can result in inconsistent energy savings compared to automated direct control systems~\cite{sardianos2020real}. The manual process can also disrupt routines and limit control over comfort, reducing overall effectiveness~\cite{popa2019deep}.

In contrast to indirect systems, direct control systems automatically adjust HVAC settings using real-time sensor data on factors like temperature, occupancy, and sometimes electricity prices. These systems autonomously reduce heating or cooling during unoccupied periods and increase output as needed to maintain setpoints~\cite{ mirakhorli2016occupancy,jazizadeh2014human,pritoni2016occupancy,chojecki2020energy}. While effective at minimizing energy consumption, it has been reported that even with the installed devices only 19.4\% of occupants actively utilize the programmable thermostats (i.e., capable of scheduled operations with a setback that relaxes the temperature set point during vacancy)~\cite{eia2018space}. Other studies indicate that despite relying on complex thermal dynamic models, 15–28\% of occupants remain dissatisfied with indoor conditions, even when they have control over the temperature setpoints~\cite{karjalainen2009thermal,amasyali2016energy}.

To address the limitations of traditional direct control systems, which rely on rigid setpoints that often fail to reflect real-time occupant preferences, more advanced forms of individual comfort preference models have been developed. In these advanced comfort models, user preferences are inferred through environmental sensors that track environmental factors such as temperature and humidity, cloth insulation, and even physiological indicators like skin temperature and heart rate~\cite{yi2015facial,ranjan2016thermalsense,jazizadeh2016can,jung2018vision,cheng2017pilot,abdallah2016sensing}. While these advanced models improve occupant comfort, they also introduce additional challenges. The deployment of sensors and the creation of accurate comfort models can be expensive, and maintaining these systems often requires users or building operators to have a certain level of expertise. Understanding how the system functions and making informed adjustments requires familiarity with the underlying technology, which can limit the broader adoption of such systems, particularly in large or diverse environments. 

Recent advancements in energy management for HVAC systems have aimed at addressing these challenges with hybrid strategies that combine the strengths of both direct and indirect control IEMS. These approaches aim to overcome the limitations of complex predefined thermal and comfort models while also avoiding the constant need for manual user intervention. \citet{shuvo2022home} develop a Home Energy Recommendation System (HERS) that applies deep reinforcement learning (RL) to manage energy consumption in smart appliances including HVAC. Their HERS dynamically learns the preferences of occupants through feedback and activity data, adjusting device operations accordingly. The system’s ability to learn and automatically adapt without relying on complex predefined models makes it flexible and efficient in providing comfort. At the same time, it can gradually reduce the burden on users to manually adjust settings. Similarly, \citet{chen2023fast} developed a HITL HVAC system using meta-learning and offline model-based RL. Their system focuses on efficiently learning occupant thermal preferences through feedback, avoiding the need for constant real-time interaction with the environment, thereby improving the practicality of applying RL to HVAC control.

These studies have provided effective solutions for learning occupant preferences and achieving energy savings without relying on complex models or requiring frequent user intervention. However, they primarily focus on optimizing internal factors like occupant comfort and energy consumption, without adequately considering external influences such as wholesale electricity price fluctuations and grid dynamics. As renewable energy sources like solar and wind increasingly penetrate the energy market, their intermittent nature introduces greater variability into the grid, driven by weather conditions and time of day~\cite{maniatis2022impact, csereklyei2019effect}. This variability leads to periods of both surplus and scarcity, resulting in fluctuations in electricity prices and supply-demand imbalances~\cite{avci2013model, amin2020optimal}. HVAC systems must, therefore, dynamically respond to these real-time changes in market conditions and energy availability to remain cost-effective and alleviate strain on the grid. Traditional systems that lack this adaptability miss opportunities to reduce operating costs during periods of excess supply and are unable to support demand response efforts critical for grid stability. Integrating real-time market dynamics into HVAC control is essential for balancing occupant comfort with economic and environmental efficiency in an increasingly renewable-driven energy landscape.

This study builds on the strengths of these previous systems to develop a real-time, adaptive HVAC control system that optimizes energy costs by incorporating both occupant comfort preferences and dynamic market signals. The proposed system aims to improve grid stability by participating in demand response efforts, and adjusting HVAC operations during periods of peak demand while maintaining high scalability and adaptability for various settings. Our main contributions can be summarized as follows. 
\begin{enumerate}
    \item We propose a simplified HITL AI framework for HVAC control that avoids the complexity of existing predefined thermal comfort models while still efficiently balancing energy cost and occupant comfort.
    \item Our system learns occupants' thermal preferences automatically via continuous adaptation to real-time feedback signals. This reduces the need for manual configuration or extensive user interaction.
    \item We incorporate real-time electricity market data into the system, enabling dynamic adjustments to HVAC operations based on fluctuating wholesale energy prices, enhancing cost efficiency and grid responsiveness.
    \item Our experimental results demonstrate that the proposed framework significantly improves energy cost savings while maintaining or enhancing occupant comfort when compared to rule-based strategies. The proposed framework also achieves performance close to the theoretical ceiling established by optimization with perfect prediction, while operating under realistic conditions with imperfect information and learned comfort preferences. 
\end{enumerate}

\section{System Model and Formulation}
\begin{figure}[t]
\includegraphics[width=\columnwidth]{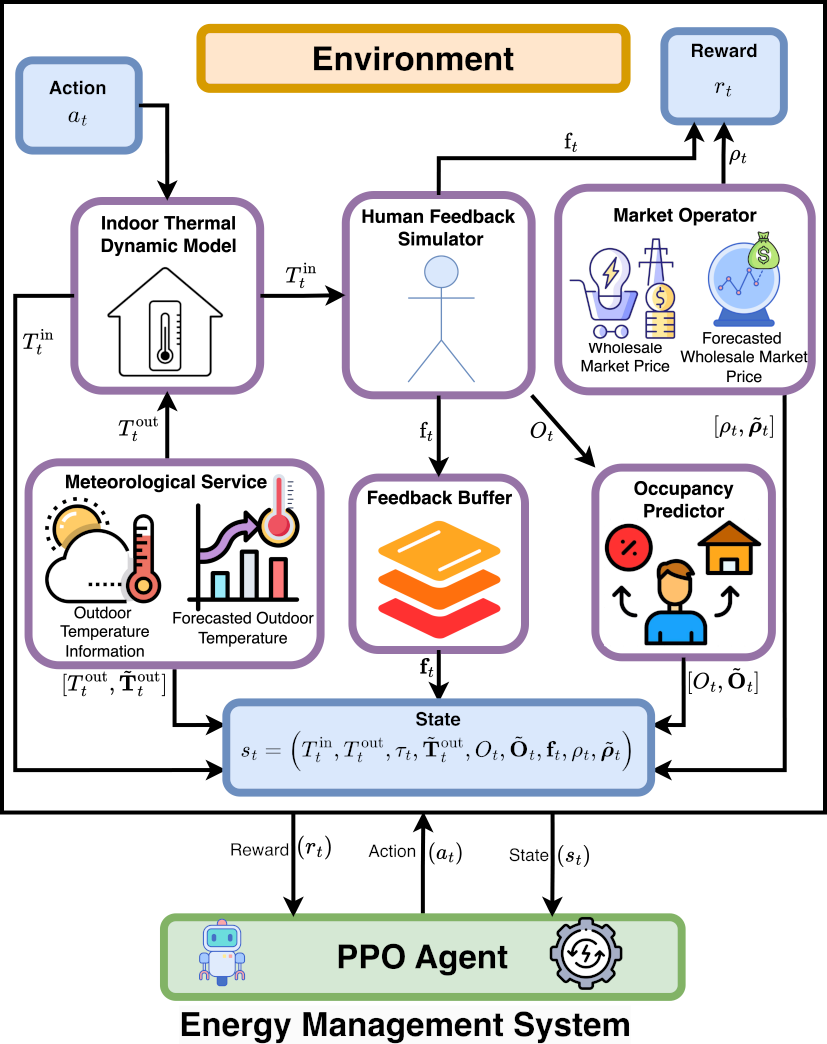}
\setlength{\belowcaptionskip}{-10pt}
\vspace{-10pt}
\Description{A plot.}
\caption{Overview of HITL Energy Management System.}
\label{fig:system-plot}
\end{figure}

This section presents our Human-in-the-Loop (HITL) system model, integrating human feedback directly into the decision-making controller. \Cref{fig:system-plot} presents a high-level overview of the system. By incorporating real-time user input alongside environmental and operational data, the system dynamically adapts to occupant preferences while balancing energy efficiency. The decision-making problem is formulated as a Markov Decision Process (MDP), which enables the system to handle uncertainties and fluctuations in occupancy and external factors, such as electricity prices. 

\subsection{MDP Model of HITL HVAC Control Problem}
The optimization of the HVAC control system is framed as a MDP that integrates a HITL approach to dynamically balance energy efficiency and occupant comfort. A key feature of this approach is its ability to learn from ongoing human feedback, enabling the system to gradually reduce the need for human intervention as it adapts to user preferences over time.

An MDP~\cite{puterman2009markov} is defined by the tuple \( (S, A, P, R, \gamma) \), where \( S \) is the state space, representing all possible states of the system, \( A \) is the action space, defining all possible actions the system can take, \( P: S \times A \rightarrow \Pr(S' \mid S, A) \) is the state transition probability function, describing the probability of transitioning from one state to another given a specific action, \( R: S \times A \rightarrow \mathbb{R} \) is the reward function, quantifying the immediate cost or benefit associated with each state-action pair and \( \gamma \in [0,1] \) is the discount factor, determining the relative importance of immediate versus future rewards.

In this HITL framework, real-time feedback from occupants is crucial for informing the system's decisions. The state of the system at any time \( t \) is represented by a state vector \( s_t \in S \) that captures the relevant environmental, operational, and feedback variables influencing the HVAC control decisions. The action \( a_t \in A \) taken at time \( t \) affects the subsequent state \( s_{t+1} \) according to the state transition dynamics \( P(s_{t+1} | s_t, a_t) \). The objective is to find an efficient policy \( \pi^* \) that maximizes the expected cumulative reward, which accounts for both energy efficiency and occupant comfort with fewer manual adjustments:
\begin{equation*}
\pi^* = \arg\max_{\pi'} \mathbb{E} \left[ \sum_{t=0}^{\infty} \gamma^t R(s_t, a_t) \Bigm\vert s_{t+1} \sim P(s_t,a_t), a_t \sim \pi'(a_t \vert s_t) \right],
\end{equation*}
where \( \pi'(a_t \mid s_t) \) is the policy that determines the probability of taking action \( a_t \) given state \( s_t \). The expectation is taken over the distribution of state-action trajectories induced by the policy \( \pi' \).

The HITL MDP framework captures the stochastic nature of the environment and system behavior, such as fluctuating outdoor temperatures, varying occupancy patterns, end-user preferences, and real-time wholesale electricity prices. The state transition dynamics are governed by the thermal properties of the space and the operational characteristics of the HVAC system, which are discussed in subsequent subsections.

\subsection{State Space Representation}
The state space \( S \) of the MDP encompasses all variables that characterize the system's environment, operational status, and occupant feedback at any given time \( t \). The state \( s_t \in S \) at time step \( t \) is represented as a vector of sensor readings that collectively describe all factors influencing the HVAC control system's decisions. 
This section describes each state variable in detail and provides the mathematical formulation that defines the state space.

\subsubsection{Indoor Temperature (\(T^\text{in}_{t}\))}
The indoor temperature, \( T^\text{in}_{t} \), represents the temperature of the controlled environment at time \( t \). It is a key variable that influences occupant comfort and can indirectly impact the system's reward through its effect on user feedback. When the indoor temperature deviates significantly from the comfort range, occupants are more likely to provide feedback that overrides the system's decision.

\subsubsection{Outdoor Temperature (\(T^\text{out}_{t}\))}
The outdoor temperature, \( T^\text{out}_{t} \), serves as an external environmental input that affects the indoor temperature, and the cooling or heating decision on the HVAC system. Fluctuations in outdoor temperature influence energy consumption, as the system must adjust its operation to counteract such conditions and maintain a comfortable indoor environment.

\subsubsection{Time of Day (\(\tau\))}
The time of day, \( \tau_t \), is a variable that captures the temporal patterns in energy usage and occupancy. Many aspects including occupancy status, outdoor temperature trends, and wholesale electricity market rate follow daily cycles. To model this periodic nature, the time of day is encoded using cyclical transformations, such as sine and cosine functions:
\begin{equation}
\tau_t = \left( \sin\left(\frac{2 \pi t}{T^\text{cycle}}\right), \cos\left(\frac{2 \pi t}{T^\text{cycle}}\right) \right),
\end{equation}
where \( T^\text{cycle} \) represents the total number of time steps in one cycle, which depends on the resolution of the data used. 
This generalized encoding allows the model to represent temporal patterns at different resolutions and anticipate changes in energy demand and occupancy that are tied to specific times within each cycle.

\subsubsection{Forecasted Outdoor Temperatures (\(\tilde{\textbf{T}}^\text{out}_{t}\))}
Forecasted outdoor temperatures, \(\tilde{\textbf{T}}^\text{out}_{t}\), provide anticipatory information about future outdoor conditions over a defined horizon. This vector contains predicted temperatures for the upcoming time steps, allowing the system to make more informed decisions about future HVAC operations. The forecasted outdoor temperature vector is defined as:
\begin{equation}
\tilde{\textbf{T}}^\text{out}_{t} = \left(\tilde{T}^\text{out}_{1}, \tilde{T}^\text{out}_{2}, \ldots, \tilde{T}^\text{out}_{h^T}\right),
\end{equation}
where \(h^T\) denotes the prediction horizon, indicating the number of future time steps over which the outdoor temperatures are forecasted. By including \(\tilde{\textbf{T}}^\text{out}_{t}\) in the state space, the system can plan actions that consider both current and anticipated future environmental conditions, thereby optimizing total energy cost.

\subsubsection{Occupancy Information (\(O_{t}\))}
This indicates whether the space is currently occupied at time \( t \). This binary variable takes values of \(1\) if the space is occupied and \(0\) otherwise. Occupancy significantly impacts both energy consumption and comfort preferences, as the HVAC system's operation is adjusted to prioritize comfort when occupants are present. The inclusion of occupancy information in the state space allows the MDP model to make context-aware decisions based on real-time occupancy data.

\subsubsection{Forecasted Occupancy Information (\(\tilde{\mathbf{O}}_{t}\))}
Forecasted occupancy information, \(\tilde{\mathbf{O}}_{t}\), represents the predicted probabilities of occupancy over future time steps, providing crucial insights for decision-making. Unlike the binary occupancy information \( O_{t} \), 
\(\tilde{\mathbf{O}}_{t}\) is generated by an occupancy predictor. This predictor forecasts the likelihood of future occupancy over a prediction horizon, allowing the system to anticipate changes in usage patterns. The forecasted occupancy vector is defined as:
\begin{equation}
\tilde{\mathbf{O}}_{t} = \left(\tilde{O}_{1}, \tilde{O}_{2}, \ldots, \tilde{O}_{h^O}\right),
\end{equation}
where \(h^O\) represents the prediction horizon for occupancy. By including \(\tilde{\mathbf{O}}_{t}\) in the state space, the model can anticipate changes in comfort and energy needs, allowing for proactive adjustments to the HVAC control strategy based on expected occupancy patterns.

\subsubsection{Feedback Buffer (\(\mathbf{f}_t\))}
The feedback buffer, \(\mathbf{f}_t\), is a vector that records instances where occupants override the system's automated decisions, providing direct feedback on their comfort preferences over a defined horizon. The feedback buffer is defined as:
\begin{equation}
\mathbf{f}_t = \left(\text{f}_{t-1}, \text{f}_{t-2}, \ldots, \text{f}_{t-h^\text{f}}\right),
\end{equation}
where \(h^\text{f}\) denotes the feedback horizon, indicating the number of past time steps for which occupant overrides are stored. Each element in the buffer, \(\text{f}_t\), can take one of three possible values: \(\text{f}_t \in \{-1, 0, 1\}\). A value of \(\text{f}_t = 1\) represents an occupant's override to turn on the AC, \(\text{f}_t = 0\) indicates no override has been made by the occupant, and \(\text{f}_t = -1\) denotes an occupant's override to turn off the AC. This formulation allows the feedback buffer to capture the direction and occurrence of overrides, thereby reflecting the occupant's comfort preferences over time.
Including \(\mathbf{f}_t\) in the state space allows the system to consider recent occupant overrides when making future decisions about HVAC operations. 
This information is crucial for preventing the system from repeatedly requiring occupant overrides, thereby improving user satisfaction and reducing manual interventions.

\subsubsection{Wholesale Market Rate (\(\rho_{t}\))}
The wholesale market rate, \( \rho_{t} \), reflects the current price of electricity in the market at time \( t \). This variable not only represents the immediate cost of energy consumption but also serves as an indicator of supply and demand dynamics on the grid. High prices may signal high demand or low supply conditions, affecting the cost of operating the HVAC system. By including the wholesale market rate in the state space, the model can optimize the AC's operation to minimize energy costs subject to grid stability and pricing signals.

\subsubsection{Forecasted Wholesale Market Rate (\(\tilde{\boldsymbol{\rho}}_{t}\))}
Forecasted wholesale market rate, \(\tilde{\boldsymbol{\rho}}_{t}\), provides predictive information about future electricity prices over a defined horizon. This vector contains the predicted market rates for upcoming time steps, which is crucial for optimizing energy usage and cost in the HVAC control strategy. The forecasted wholesale market rate vector is defined as:
\begin{equation}
\tilde{\boldsymbol{\rho}}_{t} = \left(\tilde{\rho}_{1}, \tilde{\rho}_{2}, \ldots, \tilde{\rho}_{h^\rho}\right),
\end{equation}
where \(h^\rho\) denotes the prediction horizon for the market rate. By incorporating \(\tilde{\boldsymbol{\rho}}_{t}\) into the state space, the model can plan actions that account for both current and anticipated future electricity costs, enabling more cost-effective and energy-efficient decision-making.

Given its components, the state vector at time \( t \) is defined as:
\begin{equation}
s_t = \left(T^\text{in}_{t}, T^\text{out}_{t}, \tau_t, \tilde{\mathbf{T}}^\text{out}_{t}, O_{t}, \tilde{\mathbf{O}}_{t}, \textbf{f}_t, \rho_{t}, \tilde{\boldsymbol{\rho}}_{t}\right).
\end{equation}
This formulation captures a comprehensive snapshot of the environment and system dynamics, and provides a robust basis for the system to optimize the HVAC control strategy, balancing energy efficiency, and occupant comfort.

\subsection{Action Space Definition}
Action space \(A\) consists of the possible actions our HVAC control system can take to manage the indoor environment. We assume that the heating or cooling mode is preset by the occupants, meaning the controller does not dynamically switch direction. Instead, the control focuses on scheduling when the HVAC should be turned on or off based on the preset direction. Thus, the action space is: 
\begin{equation}
A = \{0, 1\},
\end{equation}
where \(1\) represents "HVAC On," meaning the system is actively operating, and \(0\) represents "HVAC Off". 

\subsection{State Transition Dynamics}
In our MDP model for HVAC control, the state transition dynamics describe how each state variable evolves over time based on the current state and the chosen action. The transition function \(P(s_{t+1} \mid s_t, a_t)\) captures these dynamics. 
Below, we outline the transition dynamics for each state variable:

\subsubsection{Indoor Temperature Transition}
We adopt a thermal model~\cite{mortensen1988stochastic} that conditions the next indoor temperature \(T^\text{in}_{t+1}\) on three variables: current indoor temperature~\(T^\text{in}_{t}\), outdoor temperature~\(T^\text{out}_t\), and the controlled action~\(a^{c}_{t}\) at time~$t$. It is defined as
\begin{equation}
T^\text{in}_{t+1} = \alpha T^\text{in}_t + (1 - \alpha) (T^\text{out}_t + p \cdot a^{c}_{t} \cdot P),\label{eq:transition}
\end{equation}
where parameter \(P\) represents the power effect of the AC, and \(p \in \{-1,1\}\) indicates whether the HVAC is in heating or cooling mode. The parameter \(\alpha\) is a decay factor that determines how quickly the indoor temperature responds to changes, given by:
\begin{equation}
\alpha = e^{-\frac{\Delta t}{ \eta^{R} \cdot \eta^{C}}},
\end{equation}
with \(\Delta t\) being the time interval in hours, \(\eta^{R}\) being the thermal resistance of the space in \(^\circ C/\text{kW}\) and \(\eta^C\) being the thermal capacitance of the space in \(\text{kWh}/^\circ C\). 

\subsubsection{Feedback Simulation and Feedback Buffer Update}
The feedback simulation emulates how occupants might override the system due to discomfort. This process updates the feedback buffer \(\mathbf{f}_t\), which store recent simulated feedback signals. Unlike predefined setpoints used in other system, in our framework these feedback signals are crucial for guiding the agent’s decision-making.

At each time step, the feedback probability, \(p_{t}^{\text{f}}\), is calculated to determine whether an occupant would override the current system due to discomfort. This probability is influenced by the difference between the current indoor temperature, \(T_{t}^\text{in}\), and a set comfort temperature, \(T^{\text{set}}\). The feedback probability is computed as:
\begin{equation}
\textcolor{black}{
p_{t}^{\text{f}} = \min\left( \left( \frac{T_{t}^{\text{in}} - T^{\text{set}}}{\theta^\text{range}} \right)^2, \, p^\text{max} \right),
}
\end{equation}
\textcolor{black}{where \(T^{\text{set}}\) is the set comfort temperature, \(\theta^\text{range}\) is the comfort range value, and $p^\text{max}$ is the maximum probability cap that can be set to represent max level of occupant response likelihood. This likelihood accounts for scenarios where occupants might not provide feedback even under significant discomfort due to various practical constraints.}
\textcolor{black}{The quadratic relationship between temperature deviation and discomfort probability reflects the nonlinear nature of human thermal comfort perception, where discomfort increases more rapidly as temperatures deviate further from the preferred setpoint~\cite{yu2018energy,constantopoulos1991estia,bahrami2017decentralized}. Our approach extends these standard methods by linking discomfort to feedback likelihood, providing a probabilistic framework for modeling occupant response behavior.}

Once the feedback probability is determined, the new feedback, \(\text{f}_{t}\), is simulated based on whether the expected action of maintaining comfort temperature \(a_{t}^{*}\) differs from the system’s chosen action \(a_{t}\) and when the occupant is present. The expected action of maintaining comfort temperature, \(a^*\), is defined as:
\begin{equation}
a^{*}_{t} = 
\begin{cases} 
1, \quad \text{if } &\left( (T_{t}^\text{in} > T^{\text{set}} \wedge T_{t}^\text{out} > T^{\text{set}}) \right. \\
&\left. \, \vee \, (T_{t}^\text{in} < T^{\text{set}} \wedge T_{t}^\text{out} < T^{\text{set}}) \right) \\
0, & \text{otherwise.}
\end{cases}
\end{equation}

The new feedback, \(\text{f}_{t}\), is then simulated as:
\begin{equation}
\text{f}_{t} = 
\begin{cases} 
1, & \text{if } (a_{t}^{*}=1) \land (O_t = 1) \land (a_{t}^{*} \neq a_{t}) \land (X^{\text{f}}_t=1),\\
-1, & \text{if } (a_{t}^{*}=0) \land (O_t = 1) \land (a_{t}^{*} \neq a_{t}) \land (X^{\text{f}}_t=1),\\
0, & \text{otherwise.}
\end{cases}
\end{equation}

To decide if feedback is generated, we introduce the binary variable \(X^\text{f}\), which is sampled from a Bernoulli distribution, 
\begin{equation}
X^\text{f}_t \sim \text{Bernoulli}(p_{t}^{\text{f}}).
\end{equation}
Here, \(\text{f}_{t}\) takes a value of \(1\) to represent a user override to turn on the HVAC when the expected action to maintain comfort temperature, \(a_{t}^{*} = 1\), differs from the system's chosen action \(a_{t}\). This occurs when the occupant is present (\(O_t = 1\)) and the feedback condition \(X^{\text{f}} = 1\) is triggered. Conversely, \(\text{f}_{t}\) takes a value of \(-1\) to represent a user override to turn off the HVAC when the expected action is \(a_{t}^{*} = 0\) under similar conditions. If these conditions are not met (i.e., the system's action aligns with the expected action to maintain comfort, there is no occupant, or the feedback condition is not triggered) no feedback is generated, and \(\text{f}_{t} = 0\). 

After generating the new feedback value, the feedback buffer is updated. The feedback buffer, \(\mathbf{f}_t\), has a fixed length \(h^{\text{f}} = 16\), corresponding to 4 hours with each time step set to 15 minutes. At each time step, the buffer is updated by inserting the new feedback at the beginning and shifting the rest, which is defined as:
\begin{equation}
\mathbf{f}_{t+1} = [\text{f}_{t}, \text{f}_{t-1}, \text{f}_{t-2}, \ldots, \text{f}_{t+1-h^{\text{f}}}].
\end{equation}

The simulation measures the agent's ability to adapt and optimize control strategies using feedback, even with limited knowledge of occupant preferences. By incorporating simulated overrides, the agent learns to adjust actions dynamically, balancing comfort and energy efficiency. The frequency of overrides serves as a key indicator of discomfort, helping assess performance in complex, real-world scenarios with uncertain conditions.

\subsubsection{From Action to Control Decision}
After simulating and updating the feedback, the final control used in the temperature transition~\Cref{eq:transition} is determined by combining the system's action \(a_{t}\), with the new feedback value, \(\text{f}_{t}\). The final controlled action \(a^c_t\), incorporates the feedback to dynamically adjust the control strategy, ensuring that the system's decisions better align with simulated occupant preferences.
\begin{equation}
a^c_t = 
\begin{cases} 
1, & \text{if } (a_{t} = 1 \wedge \text{f}_{t} = 0) \vee (a_{t} = 0 \wedge \text{f}_{t} = 1), \\
0, & \text{otherwise.}
\end{cases}
\end{equation}

\textcolor{black}{The feedback mechanism operates through two distinct processes: immediate environmental control and policy adaptation. When occupants provide feedback through overrides, it immediately affects the HVAC operation within the current timestep, similar to direct thermostat adjustments in real-world scenarios. However, the learning and adaptation of the control policy occurs through the feedback buffer, which incorporates these signals into the decision-making process from timestep $t+1$ onwards. This separation ensures responsive environmental control while maintaining structured policy learning through the feedback buffer mechanism.}

This logic enables the system to integrate simulated feedback and dynamically adapt to occupant discomfort while improving energy efficiency and comfort. By responding to feedback, the model ensures actions remain flexible and accommodate changes in preferences.

\subsubsection{State Variables Based on External Data and Internal Predictors}

Our MDP model's state variables are updated from both external data sources and internal predictors. External variables, such as occupancy information (\(O_t\)) from sensors or GPS signals, outdoor temperature (\(T^\text{out}_t\)), forecast outdoor temperature (\(\tilde{\mathbf{T}}^\text{out}_{t}\)), wholesale market rates (\(R_t\)), and forecast market rates (\(\tilde{\mathbf{R}}_{t}\)), can be sourced from established services like weather forecasts and market predictions. These reliable, real-time inputs allow the system to optimize energy use and maintain comfort with minimal complexity.

Other variables, such as forecasted occupancy information (\(\tilde{\mathbf{O}}_{t}\)), require internal prediction due to the unique occupancy patterns specific to each household. The internal predictor uses historical data and time of day as contextual factors to estimate future occupancy probabilities over a given horizon. This customized approach captures occupant behavior, allowing the system to make proactive decisions—such as pre-heating or pre-cooling—where future occupancy is uncertain and dynamic.

\subsection{Reward Design}
The reward function, \(R(s, a, s')\) in the proposed HVAC system leverages the HITL approach, where real-time occupant feedback directly informs operational decisions. Unlike conventional approaches with static setpoints, this reward structure dynamically integrates user input to calculate discomfort cost. Occupant interventions and its frequency, are reflected in the discomfort cost, while energy costs are tied to real-time electricity prices. This approach enables the system to adapt to user preferences, reducing the need and frequency for manual intervention over time while optimizing both comfort and energy efficiency.

\subsubsection{Discomfort Cost Calculation}
The discomfort cost component of the reward is determined by the feedback received from occupants regarding their comfort levels. Unlike traditional approaches that rely on predefined temperature setpoints and measure deviations from these setpoints, our model dynamically adjusts based on the feedback that directly reflects occupant comfort or discomfort. The discomfort cost, \(C^{\text{discomfort}}\), is calculated using a weighted sum of the feedback buffer, which stores recent feedback values:
\begin{equation}
C^{\text{discomfort}}_t = 
\begin{cases} 
{\sum_{i=1}^{h^{\text{f}}} w^{\text{discomfort}}_i \cdot |\text{f}_{t-i}|}, & \text{if } (\text{f}_{t} = 1) \vee (\text{f}_{t} = -1), \\
-\epsilon, & \text{if } (\text{f}_{t} = 0) \land (O_{t} = 1),\\
0, & \text{otherwise,}
\end{cases}
\end{equation}
where \(w^{\text{discomfort}}_i\) is the weight for feedback entry of each time step in the feedback buffer formulated as:
\begin{equation}
    w^{\text{discomfort}}_i = e-e^{\frac{i}{h^{\text{f}}}}.
\end{equation}

The discomfort weights \(w^{\text{discomfort}}_i\) are designed such that more recent feedback receives a heavier penalty. This weighting scheme discourages frequent user interventions, as high intervention frequency indicates worse discomfort and a failure of the system to maintain satisfactory conditions. By penalizing more recent feedback more heavily, the model is incentivized to learn and adapt its control strategy to prevent repeated overrides by occupants. If the space is occupied (\(O_{t} = 1\)) but no feedback is received (\(\text{f}_{t} = 0\)), a small negative value, \(-\epsilon\), is applied. This encourages the model to continue with its current action, as the absence of feedback suggests that no immediate change is necessary. This approach promotes stability and prevents unnecessary adjustments that could lead to discomfort or inefficient energy use. Otherwise, when the space is unoccupied the discomfort cost is set to zero, reflecting that maintaining comfort is not a priority without occupants present. This dynamic approach to discomfort cost calculation ensures that the agent learns to balance energy efficiency and occupant comfort effectively by directly responding to occupant feedback and minimizing high-frequency interventions.

\subsubsection{Energy Cost Calculation}
Energy costs~\(C_{\text{energy}}\) are a function of the power consumption of the HVAC unit and the wholesale market price. It is calculated as:
\begin{equation}
C^{\text{energy}}_t = a^c_t \cdot P^{\text{HVAC}} \cdot \Delta t \cdot \rho_t,
\end{equation}
where we set the power consumption of the HVAC when operating \(P^{\text{HVAC}} = 3.5\text{kW}\), \(\Delta t = 0.25 \, \text{hours}\) (i.e., a 15-minute time step), and \(\rho_t\) is the current wholesale market rate for electricity. This formulation ensures that the model considers both the operational costs of running the HVAC and the varying electricity prices throughout the day, reflecting real-world economic factors.

\subsubsection{Total Cost and Reward Calculation}

The total cost, \(C^{\text{total}}_t\), combines both discomfort and energy costs, weighted by a parameter \(\beta\) that balances the importance of comfort versus energy efficiency:
\begin{equation}
C^{\text{total}}_t = \beta \times C^{\text{discomfort}}_t + (1 - \beta) \times C^{\text{energy}}_t,
\end{equation}
where \(\beta\) is a tunable parameter that determines the relative weight of discomfort in the total cost calculation. A higher \(\beta\) places more emphasis on minimizing discomfort, while a lower \(\beta\) prioritizes energy cost savings.

The reward is then calculated as the negative of the total cost:
\begin{equation}
r_t = -C^{\text{total}}_t.
\end{equation}

\section{Solution Methodology}
This section outlines the approach used to develop and optimize control strategies for balancing energy efficiency and occupant comfort within the MDP framework. The proposed methodology integrates a predictive model for forecasting occupancy information and applies a RL-based optimization technique to learn effective control policies. First, we describe the integration of the occupancy prediction model, which generates future occupancy states based on historical data and contextual factors. This is followed by an explanation of the Proximal Policy Optimization (PPO) algorithm~\cite{schulman2017proximal}, which is employed to optimize the control policies. The PPO algorithm enables the agent to make decisions that minimize energy consumption while responding dynamically to occupant feedback to maintain comfort. Together, these components provide a comprehensive framework for managing energy use in dynamic environments with uncertain occupancy patterns.

\subsection{Occupancy Prediction Model Integration}
Accurate prediction of future occupancy is essential for optimizing control strategies within the MDP framework, as occupancy patterns significantly affect both energy consumption and comfort levels. Unlike weather and market data, which can be reliably obtained from external forecasting services, occupancy information is highly specific to each building and its occupants. Thus, an internal prediction model is developed to estimate future occupancy probabilities based on historical data and temporal information at each time step \(t\).

The proposed occupancy prediction model employs a two-layer bidirectional Long Short-Term Memory (LSTM) network to capture both short-term and long-term dependencies in occupancy patterns. At each time step \(t\), the model utilizes the past occupancy information up to and including the current time step. The past occupancy vector is denoted as:
\begin{equation}
\mathbf{O}^{\text{p}}_t = [O_{t-h^{\text{p}}}, O_{t-h^{\text{p}}+1}, \ldots, O_{t}],
\end{equation}
where \(h^{\text{p}}\) represents the horizon of past information, and each element \(O_{t-i}\) for \(i \in \{ 0, 1, \ldots, h^{\text{p}}\}\) is the observed occupancy status at time step \(t-i\).

Similarly, the temporal feature vector for the past horizon is defined as:
\begin{equation}
\boldsymbol{\tau}^{\text{p}}_t = [\tau_{t-h^{\text{p}}}, \tau_{t-h^{\text{p}}+1}, \ldots, \tau_{t}],
\end{equation}
where each \(\tau_{t-i}\) captures the periodic nature of time through sine and cosine transformations. 

The concatenated sequence \([\mathbf{O}_{t-i}, \boldsymbol{\tau}_{t-i}]\) for each time step is processed by the first bidirectional LSTM layer, generating hidden states in both forward and backward directions. The forward hidden state at time step \(t-i\) is denoted as \(\overrightarrow{h}^{\text{p}}_{t-i}\) and the backward hidden state as \(\overleftarrow{h}^{\text{p}}_{t-i}\). The past occupancy context vector \(\textbf{H}^{\text{p}}_t\) is constructed by concatenating the final forward hidden state from the current time step and the backward hidden state from the earliest time step within the horizon:
\begin{equation}
\textbf{H}^{\text{p}}_t = [\overrightarrow{h}^{\text{p}}_{t}; \overleftarrow{h}^{\text{p}}_{t-h^{\text{p}}}].
\end{equation}

To predict occupancy for multiple future time steps, we define the prediction horizon as \(h^{{O}}\) and \(j \in \{1, 2, \ldots, h^{{O}}\}\). For each future time step \(t+j\), the input to the second bidirectional LSTM layer is constructed by combining the context vector \(H^{\text{p}}_t\) with the corresponding temporal features:
\begin{equation}
\textbf{Z}_{t+j} = [\textbf{H}^{\text{p}}_t; \tau_{t+j}].
\end{equation}

The second LSTM layer processes these combined inputs \(\textbf{Z}_{t+j}\) to generate a hidden representation \(\textbf{h}_{t+j}^{\text{f}}\) for each future time step:
\begin{equation}
\textbf{h}_{t+j}^{\text{f}} = [\overrightarrow{h}_{t+j}^{\text{f}}; \overleftarrow{h}_{t+j}^{\text{f}}],
\end{equation}
where \(\overrightarrow{h}_{t+j}^{\text{f}}\) and \(\overleftarrow{h}_{t+j}^{\text{f}}\) are the forward and backward hidden states of the second LSTM layer.

The predicted occupancy probability \(\tilde{O}_{t+j}\) for each future time step \(t+j\) is then obtained by passing the hidden representation \(h_{t+j}^{\text{f}}\) through a fully connected layer followed by a sigmoid activation:
\begin{equation}
\tilde{O}_{t+j} = \sigma(W^{\text{f}} h_{t+j}^{\text{f}} + b^{\text{f}}),
\end{equation}
where \(W^{\text{f}}\) and \(b^{\text{f}}\) are the weights and biases of the fully connected layer, and \(\sigma(\cdot)\) represents the sigmoid activation function.

These predicted occupancy probabilities \(\tilde{O}_{j}\) for the future time steps within the horizon \(h^{\text{O}}\) are integrated into the state space. By incorporating these probabilities, the RL agent can anticipate future states and dynamically optimize control actions based on the predicted occupancy and the cost of energy. When spaces are predicted to be unoccupied, the agent may decide to perform preheating or precooling actions if it is more cost-effective. This approach allows the agent to manage the environment in a way that prepares for future occupancy while optimizing energy costs based on forecast prices and demand.

\subsection{PPO-based Learning Control Policy}
We use the Proximal Policy Optimization~\cite[PPO]{schulman2017proximal} algorithm to learn a control policy that optimises energy efficiency and occupant comfort. By interacting with a simulator providing real-time state information such as occupancy, energy market rate, and environmental factors, PPO learns adaptive control strategies that dynamically adjust energy usage. This includes deciding when to activate or deactivate HVAC systems to perform preheating or precooling actions. PPO facilitates robust learning in these dynamic environments by updating control policies iteratively, while preventing overly large updates that could destabilize the learning process. This allows the system to handle uncertainties effectively, saving energy costs without compromising occupant comfort.

\subsubsection{Policy Representation and Objective Function}
The control policy \(\pi'_{\theta}(a_t | s_t)\), parameterized by \(\theta\), defines the probability of selecting an action \(a_t\) given the current state \(s_t\). The goal is to maximize the expected cumulative reward over time, reflecting cost-effective energy consumption while maintaining comfort. PPO optimizes this objective by improving the policy iteratively using a clipped surrogate objective function \(L^{\text{policy}}(\theta)\) to prevent overly large updates and stablize learning:
\begin{equation}
L^{\text{policy}}(\theta) = \mathbb{E}_t \left[ \min\left( r_t(\theta) \cdot \hat{A}_t, \text{clip}(r_t(\theta), 1 - \epsilon, 1 + \epsilon) \cdot \hat{A}_t \right) \right],
\end{equation}
where \(\hat{A}_t\) is the advantage function, \(\epsilon\) is a clipping hyperparameter and \(r_t(\theta)\) is the probability ratio between the updated policy and the old policy formulated as:
\begin{equation}
     r_t(\theta) = \frac{\pi'_{\theta}(a_t | s_t)}{\pi'_{\theta_{\text{old}}}(a_t | s_t)}.
\end{equation}

\subsubsection{Advantage Estimation and Value Function Approximation}
To guide the policy updates in PPO, advantage estimation is used to determine how favorable an action \(a_t\) is compared to the average action taken from a state \(s_t\). The Generalized Advantage Estimation (GAE) method is utilized to compute the advantage function \(\hat{A}_t\), which provides a more stable and less noisy estimate than the simple temporal difference (TD) error. GAE combines multi-step returns in a recursive manner to balance bias and variance effectively, defined as:
\begin{equation}
\hat{A}_t = \delta_t + (\gamma \lambda) \hat{A}_{t+1},
\end{equation}
where \(\delta_t = r_t + \gamma V(s_{t+1}) - V(s_t),
\) is the TD error, \(\gamma \in [0, 1]\) is the discount factor, and \(\lambda \in [0, 1]\) is the parameter that controls the bias-variance trade-off.

The value function \(V(s_t)\), parameterized by \(\phi\) is learned by minimizing the mean squared error between the predicted value and the actual return. The loss function is defined as:
\begin{equation}
L^{\text{VF}}(\phi) = \mathbb{E}_t \left[ \sum_{t} \left( V_{\phi}(s_t) - R_t \right)^2 \right],
\end{equation}
where \(V_{\phi}(s_t)\) is the critic estimate at \(s_t\), and \(R_t\) is the actual return.

\subsubsection{Action Space and Exploration Strategy}
The action space includes various control actions such as performing preheating or precooling, or switching systems on or off. To prevent the algorithm from converging prematurely to suboptimal strategies, an entropy bonus is added to the objective function to maintain an balance between exploration and exploitation:
\begin{equation}
L^{\text{entropy}}(\theta) = - \mathbb{E}_t \left[ \sum_{a} \pi'_{\theta}(a | s_t) \log \pi'_{\theta}(a | s_t) \right].
\end{equation}

\subsubsection{Optimization and Training Strategy}
The final objective function used for PPO consists of three main components: the policy loss, the value function loss, and the entropy bonus. The combined loss function is given by:
\begin{equation}
L(\theta, \phi) = L^{\text{policy}}(\theta) + c_1 L^{\text{VF}}(\phi) + c_2 L^{\text{entropy}}(\theta),
\end{equation}
where \(c_1\) and \(c_2\) are coefficients that balance the contributions of the value function loss and entropy bonus relative to the policy loss.
By minimizing this final combined loss ensures that the algorithm not only improves its control policy but also maintains accurate value function predictions and explores the action space sufficiently to avoid suboptimal solutions.

\section{Experiment Setting}
\subsection{Data Description and Preprocessing}
In this study, we utilize multiple data sources to optimize indoor HVAC control, focusing on factors such as occupancy information, outdoor temperature, and wholesale electricity market rates. 

The occupancy data used in our experiment is derived from the ARAS human activity recognition dataset~\cite{alemdar2013aras}. The data we used provide 30 days of measurements collected from a real home with multiple residents. The data were captured at a 1-second resolution using 20 ambient binary sensors, including force-sensitive resistors, pressure mats, contact sensors, and temperature sensors, which monitored 27 distinct activities. In our study, we focused particularly on the "Going Out" activity label, as it was crucial for accurately tracking the occupancy status of each resident. To integrate this occupancy data into our HVAC control system, we performed preprocessing steps to differentiate the occupancy status of individual residents, and then aggregated the data to reflect overall house as unoccupied only when all residents were simultaneously away. This approach ensures that the HVAC system responds appropriately to the actual occupancy status, thereby improving both energy efficiency and comfort.

We sourced outdoor temperature data from the Visual Crossing~\cite{visualcrossing_weather} weather data service. The data were recorded on an hourly basis for the location Clayton, Melbourne, VIC 3168, Australia. We utilized the temperature data starting from May 1, 2023, to May 30, 2023, to represent the outdoor environmental conditions. This data plays a critical role in adjusting the HVAC system's operations to maintain indoor comfort levels in response to changing outdoor temperatures.

The wholesale electricity market price data were obtained from the Australian Energy Market Operator (AEMO) and are accessible via AEMO Aggregated Price and Demand Data~\cite{aemo_aggregated_data}. We extracted aggregated price and demand data for Victoria, also from May 1, 2023, to May 30, 2023, ensuring the location and time information aligns with the outdoor temperature data.

To ensure consistency across all data sources, we resampled all data into a 15-minute resolution. We split our 30-day dataset into 23 training and 7 test days.

\subsection{Scenarios for HITL Policy}
In our research, we designed four distinct scenarios to evaluate how different types of occupancy information and its forecasting impact the performance of our HITL policy in controlling HVAC systems. For all scenarios, we set the comfort temperature \(T^{\text{set}} = 22^\circ C\) and comfort range \(\theta^\text{range} = 3\). \textcolor{black}{This comfort range of $\pm 3^\circ C$ is based on the acceptable thermal comfort zone for office environments specified in ASHRAE Standard 55~\cite{ashrae2020standard}, ensuring our system maintains balanced occupant comfort and energy efficiency. The $22^\circ C$ setpoint temperature is informed by the meta analysis in ~\cite{seppanen2006room} which found productivity peaks at this level, allowing our system to provide a comfortable temperature for occupants.} \textcolor{black}{The RC time constant $\eta^{R} \cdot \eta^{C}$ in Equation (9) is calculated following \cite{mortensen1988stochastic} using observable heating and cooling behavior:
\begin{equation}
\eta^{R} \cdot \eta^{C} = -\frac{1}{3600 \cdot \ln\left(\left(\frac{T_{lower}}{T_{upper}}\right)^{\frac{1}{t_{cool}}}\right)}
\end{equation}
where $T_{lower} = 21.5^{\circ}\text{C}$, $T_{upper} = 22.5^{\circ}\text{C}$, and $t_{cool} = 2700 \text{ seconds}$ is the time to cool from $T_{upper}$ to $T_{lower}$ when the system is turned off and the outdoor temperature is $0^{\circ}\text{C}$. This results in $\eta^{R} \cdot \eta^{C} = 16.50 \text{ hours}$, which aligns with typical residential building time constants ranging from 15 to 55 hours observed in \cite{john2018estimating}.}

These scenarios are intended to compare the availability and utility of occupancy data, which is often difficult to obtain and sensitive for users to share, yet crucial for optimizing energy efficiency. Across all scenarios, the agent's state consistently includes time of day, current indoor and outdoor temperatures, forecasted outdoor temperature, and both current and forecasted wholesale market rates. These elements are included because they are either easily collected internally or reliably sourced externally. Additionally, we tested 2 hours prediction horizons in each scenario to comprehensively analyze the impact of forecast lengths on the agent's decision-making process. \textcolor{black}{This horizon length is determined through empirical evaluation, balancing prediction accuracy with computational efficiency. While longer horizons were considered, they provided diminishing returns in terms of performance improvement while significantly increasing the state space complexity and training overhead. The 2-hour horizon proved sufficient for capturing relevant temporal patterns while maintaining system responsiveness, as the RL agent learns to implicitly model longer-term dynamics through training. Additionally, this horizon aligns with the thermal characteristics of typical residential buildings, where the impact of control decisions becomes increasingly uncertain beyond the 2-hour timeframe due to varying occupancy patterns and environmental conditions.} The following scenarios differ primarily in how they handle occupancy information:

\subsubsection{Scenario 1 (S1): Perfect Prediction Scenario}
This scenario includes actual current and future data for occupancy across multiple forecast horizons in the state of the HITL agent. It represents the most ideal scenario, allowing the HITL agent to make decisions with perfect future insights. This scenario serves as a benchmark to evaluate the effectiveness of occupancy information on the overall system performance.

\subsubsection{Scenario 2 (S2): No Occupancy Prediction Scenario}
In this scenario, occupancy information is entirely excluded from the agent's state. This setup tests whether the HITL agent can compensate for the lack of explicit occupancy data by solely relying on past feedback, which can implicitly reflect occupancy patterns. The feedback buffer serves as a hidden representation of occupancy, as feedback is only provided when occupants are present and uncomfortable. This scenario helps determine if the system can operate efficiently without direct occupancy inputs.

\subsubsection{Scenario 3 (S3): Current Occupancy Only Scenario}
Here, the scenario includes occupancy information limited to the current time step in the agent's state. This setup tests how the availability of only immediate occupancy data affects the agent’s decision-making and performance, compared to having full occupancy forecasts (S1) or none at all (S2). This comparison provides insights into the trade-offs between using current versus predicted occupancy information.

\subsubsection{Scenario 4 (S4): Realistic Predicted Occupancy Scenario}
This scenario integrates current occupancy information and predicted occupancy probability based on time of the day information and past occupancy information. It is designed to reflect real-life situations where occupancy information must be forecasted rather than known with certainty. Comparing this scenario to the Perfect Prediction Scenario (S1) helps assess the impact of prediction accuracy on system performance. Additionally, comparing this scenario to the Current Occupancy Only Scenario (S3) illustrates how adding predicted occupancy data can influence overall performance.

\subsection{\textcolor{black}{Sensitivity Analysis Settings}}

\textcolor{black}{
To evaluate the robustness of our control scheme against imperfect human feedback behavior, we examined system performance under different maximum probability caps $p^\text{max}$. The RL agent is first trained with $p^\text{max} = 1.0$, representing an ideal scenario where occupants always provide feedback when experiencing discomfort. During the evaluation, we tested different values of $p^\text{max}$ (from 0.50 to 1) to simulate more realistic scenarios where occupants might not provide feedback even under significant discomfort. This training-evaluation approach allows us to assess whether a policy learned under ideal conditions can maintain performance when faced with more uncertain human feedback patterns.
}

\subsection{Benchmark Controllers}
To further evaluate the performance of our HITL HVAC control policy, we compare it against two benchmark policies: a rule-base controller and an optimization controller.

\subsubsection{Rule-base Controller}
In this controller, the HVAC system mimics a common approach: the HVAC is turned on when people are home and turned off when they leave. The system operates based on a preset temperature set point, maintaining the indoor temperature when the house is occupied and shutting down when the house is unoccupied. This reflects a everyday practice where the HVAC system is not responsive to dynamic pricing signals but instead operates purely based on the current occupancy status.

\subsubsection{Optimization Controller}
The HVAC control in this controller is formulated as an optimization problem, solved using the Gurobi solver with a rolling horizon approach. The optimization problem utilizes the same thermal dynamic model as employed in our RL environment, ensuring consistency in the thermal behavior of the environment and the HVAC system. This controller assumes perfect predictions of future outdoor temperatures, wholesale market prices, and occupancy status. Hard constraints are set to maintain the indoor temperature within a specified comfort range. \textcolor{black}{We tested various ±3-degrees Celsius comfort ranges and 24 hours prediction horizon to ensure it has the same comfort settings with our HITL policy and explore its energy costs. We use the perfect prediction of outdoor temperature, wholesale market rate and occupancy status in this controller, and long prediction horizon to reflect the near-optimal system performance under ideal conditions.}

\section{Results and Analysis}
\begin{figure*}[ht!]
\subfigure[\textcolor{black}{Cost analysis of HVAC control strategies across different values of $\beta$ with 2 hours prediction horizon, where $\beta$ is the weighting parameter that determines the relative importance of discomfort versus energy costs as defined in Equation (19). Results are shown with maximum feedback probability cap $p^{\text{max}} = 1$.}\label{fig:cost_analysis}]{\includegraphics[width=\textwidth]{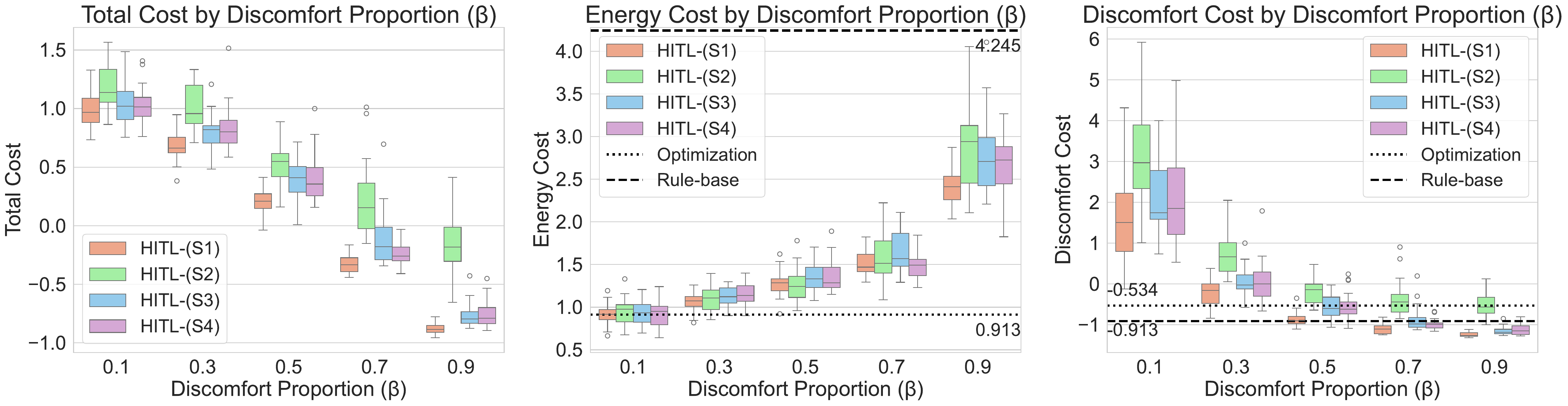}}\\
\subfigure[Temperature trajectories over time.\label{fig:temperature_analysis}]{\includegraphics[width=1.03\columnwidth]{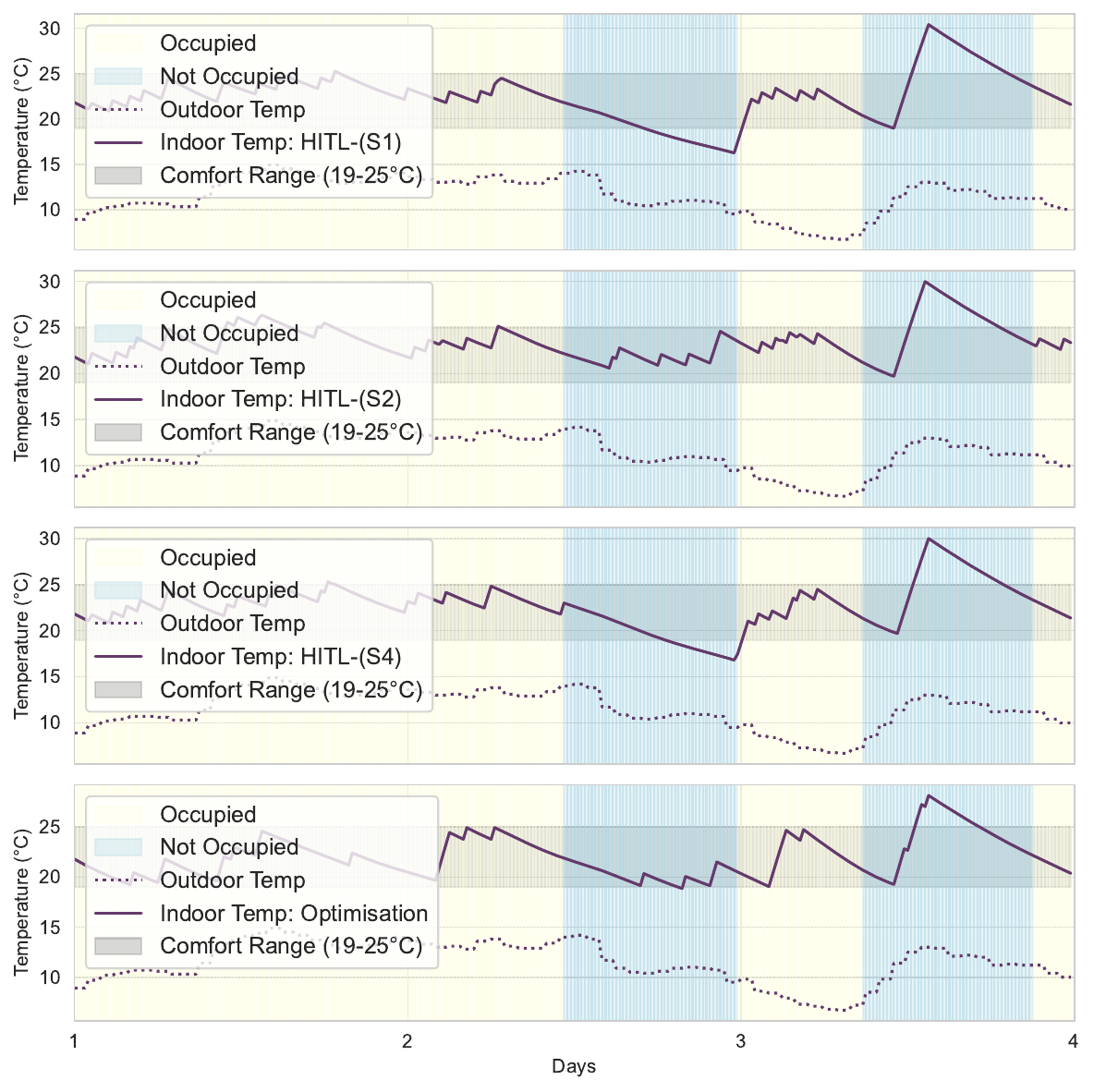}}
\hfill
\subfigure[Controller decisions and user feedback signals over time.\label{fig:system_decision_feedback}]{\includegraphics[width=1.03\columnwidth]{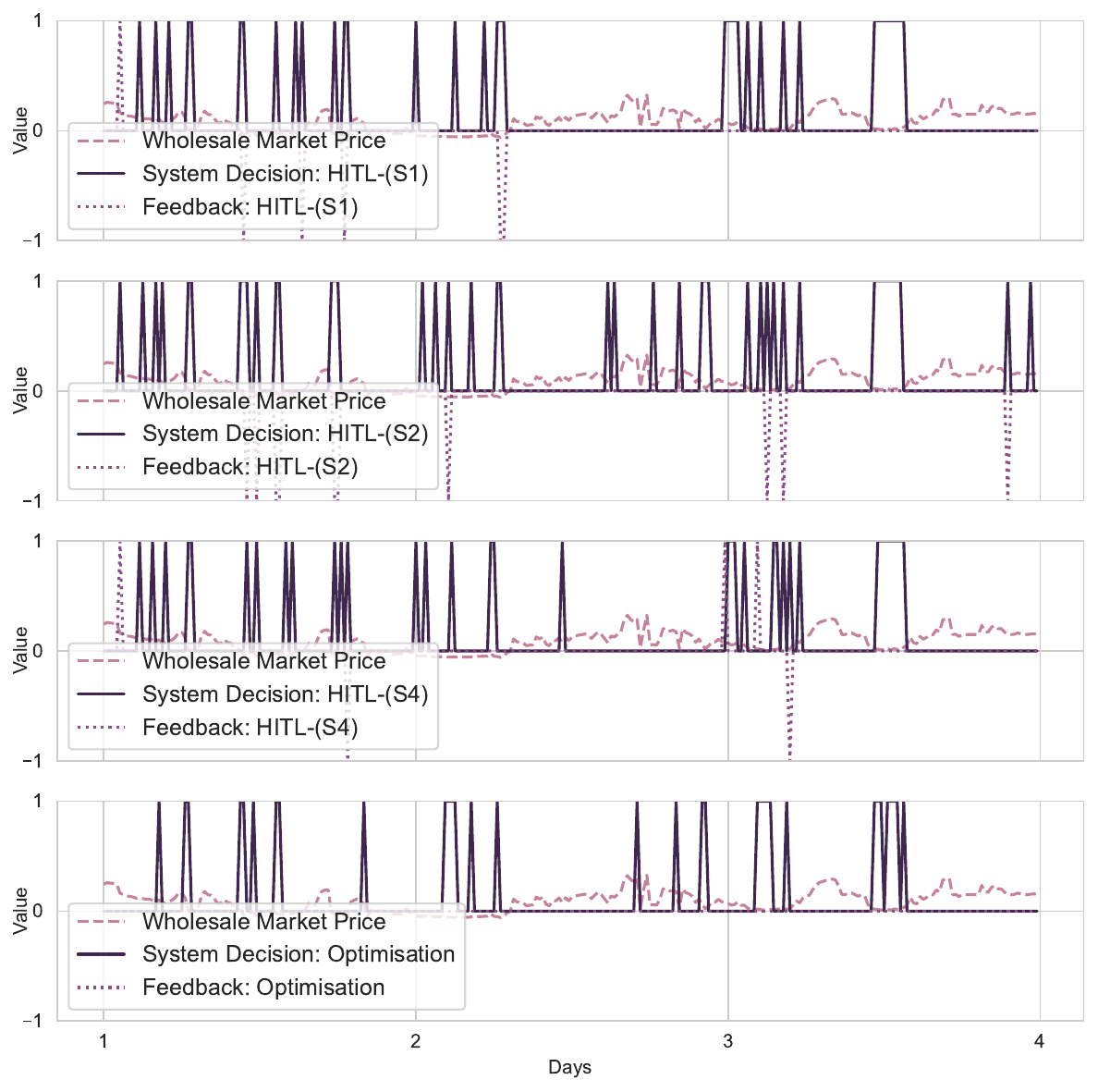}
}
\Description{A plot.}
\vspace{-10pt} 
\setlength{\belowcaptionskip}{-10pt}
\caption{Comparison of different control strategies in terms of objectives, temperatures, and control decisions plus feedback.}\label{fig:full_comparison}
\end{figure*}

This section presents an in-depth analysis of the proposed HITL HVAC control framework, focusing on its performance across different occupancy scenarios compared to conventional control systems and its robustness to imprecise human feedback. The evaluation considers key metrics such as energy cost, occupant comfort, indoor temperature maintenance, and system responsiveness, all illustrated in Figures \ref{fig:full_comparison} and \ref{fig:sensitivity_analysis}. \textcolor{black}{Furthermore, we evaluate temperature control performance using two key metrics: the temperature violation probability (i.e. percentage of time the temperature falls outside the comfort range during occupied periods) and Mean Absolute Error (MAE) to setpoint temperature (i.e. average absolute difference between indoor temperature and the desired setpoint of 22°C when the space is occupied). These metrics provide additional insights into how well each approach maintains desired comfort conditions.}

\textcolor{black}{Table \ref{tab:metric_comparison} presents key temperature performance metrics comparing our HITL approach with the optimization and rule-based methods. For the HITL approach, these metrics are calculated by averaging results across 25 independent runs for all scenarios and discomfort proportions to provide a comprehensive evaluation.} Figure \ref{fig:cost_analysis} presents a box plot that illustrates the total cost, energy cost, and discomfort cost across the four HITL scenarios. The box plot contains data from 25 independent runs of the RL algorithm, capturing the variability in the results due to the stochastic nature of the learning process. The figure also incorporates various discomfort proportions, which represent how much weight the system places on occupant discomfort versus energy cost. By adjusting these proportions, the system can prioritize either energy savings or occupant comfort, providing insight into the trade-offs made in each scenario. \textcolor{black}{Additionally, the figure includes energy cost and discomfort cost for both rule-based and optimization-based methods with 24-hour perfect prediction horizon. For fair comparison, these benchmark methods' performances are evaluated within the same RL environment over 25 independent runs, with the median values selected as final results.} To further explore the system’s behavior, Figures \ref{fig:temperature_analysis} and \ref{fig:system_decision_feedback} focus on temperature maintenance and system decision-making in response to feedback. For these figures, Scenario 3 (current occupancy only) and the rule-based algorithm are excluded from the comparison due to page limitations and the view that these cases are not representative enough for these particular analysis. Instead, S1 (Perfect Prediction), S2 (No Occupancy Information), S4 (Realistic Occupancy Prediction), and the optimization controller are included for comparison. These scenarios allow us to explore the impact of prediction accuracy, occupancy information, and dynamic system adaptation.

The following subsections will delve deeper into these results, analyzing the implications of each scenario and providing a detailed discussion on how the HITL framework adapts to real-time conditions and balances the competing demands of energy efficiency and occupant comfort.

\subsection{Learning Preferences from Real-Time Feedback}
The integration of a real-time feedback mechanism allows the system to effectively learn occupant comfort preferences, requiring minimal manual overrides. In S1 (perfect prediction) and S4 (realistic prediction), the system maintains indoor temperatures within the desired comfort range when the room is occupied, with limited need for user intervention, as demonstrated in Figure \ref{fig:temperature_analysis} and \ref{fig:system_decision_feedback}. This suggests that the feedback mechanism adapts to occupant needs over time, ensuring the system remains responsive while reducing the frequency of manual overrides. Even in S2, where no explicit occupancy information is provided, the system is still able to maintain reasonable performance by relying on historical feedback. This is reflected in Figure \ref{fig:system_decision_feedback}, where S2 achieves a reasonable number of overrides despite the absence of occupancy data. These results suggest that the feedback mechanism itself can play a key role in maintaining comfort and energy efficiency, even in the absence of real-time occupancy information.

\subsection{Impact of Occupancy Prediction Accuracy}
The incorporation of occupancy information can significantly enhance system performance, even with the feedback mechanism, as indicated by the comparison of energy and discomfort costs across different scenarios shown in Figure \ref{fig:cost_analysis}.
While S2 performs moderately well in terms of energy costs despite lacking occupancy data, S1, S3, and S4 demonstrate marked improvements when occupancy data (i.e., whether real-time or predicted) is included in the system’s decision-making process.

The impact of occupancy prediction becomes evident when considering different discomfort proportions, which reflect the system’s prioritization of occupant comfort over energy savings. S1, with perfect occupancy prediction, consistently outperforms both S3 (no prediction) and S4 (realistic prediction) across all discomfort proportions, as shown in figure \ref{fig:cost_analysis}. When the discomfort proportion is set to 0.1, 0.3, and 0.5, S3 and S4 achieve similar total costs, indicating that when occupant comfort is less prioritized, realistic occupancy prediction (S4) does not provide a significant advantage over no prediction (S3). This suggests that if the prediction accuracy is not sufficiently high, it may not lead to better outcomes. However, as the discomfort proportion increases, S4 starts to outperform S3, reflecting the value of prediction accuracy in managing comfort. Figure \ref{fig:cost_analysis} shows that with 92.52\% prediction accuracy, S4 is able to better anticipate occupancy changes and maintain comfort under higher discomfort weights, whereas S3 struggles. This illustrates that while perfect prediction (S1) always leads to better outcomes, reasonably accurate predictions (S4) can still offer substantial benefits when comfort is prioritized.

\subsection{\textcolor{black}{Comparing HITL RL with Optimisation}}
\textcolor{black}{The optimization-based control method, leveraging perfect prediction and explicit comfort constraints, demonstrates strong performance in temperature maintenance as shown in Table \ref{tab:metric_comparison}. However, this performance comes with important caveats. First, it assumes perfect prediction of future states, including occupancy patterns, weather conditions, and market prices - conditions impossible to achieve in real-world applications. Second, it operates with explicit knowledge of comfort preferences, while our HITL approach must learn these preferences through interaction. As seen in Figure \ref{fig:cost_analysis}, the energy costs across different control strategies reveal interesting patterns. The optimization-based controller achieves efficient energy management under ideal conditions. The HITL approach, while operating under more realistic constraints of imperfect information and learned preferences, maintains competitive performance in balancing comfort and energy efficiency. This demonstrates the practical viability of our approach in real-world settings where perfect prediction and explicit comfort preferences are unavailable. This is also evident by its ability to adapt to changing conditions and learn user preferences over time, as shown in Figures \ref{fig:temperature_analysis} and \ref{fig:system_decision_feedback}. While the optimization-based controller relies on perfect future knowledge for decision-making, our approach demonstrates robust performance through adaptation to feedback and market conditions.}
\begin{table}[bt]
\centering
\setlength{\belowcaptionskip}{-10pt}
\caption{\textcolor{black}{Temperature Performance Metrics Comparison}}
\vspace{-10pt} 
\begin{tabular}{lrr}
\toprule
    \textbf{Method} & \textbf{Violation Probability} & \textbf{MAE to Setpoint} \\
    \midrule
    HITL & 10.24\% & 1.82 \\
    Optimization & 0.00\% & 1.60 \\
    Rule-based & 3.13\% & 0.49 \\
\bottomrule
\end{tabular}
\label{tab:metric_comparison}
\end{table}

\subsection{\textcolor{black}{Sensitivity to Feedback Responsiveness}}
\begin{figure}[tb]
    \centering
    \includegraphics[width=\columnwidth]{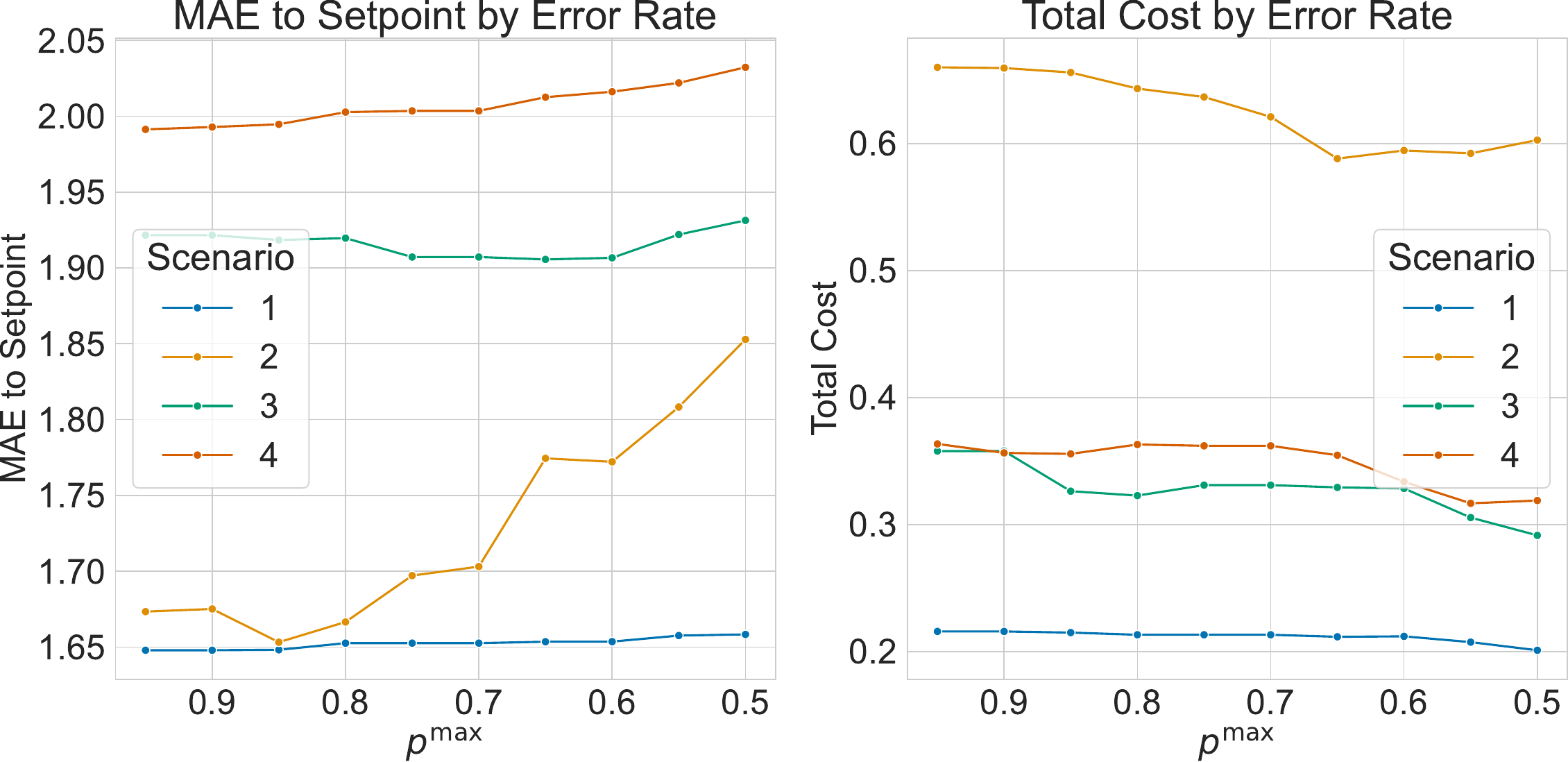}
    \vspace{-10pt}
    \Description{A line graph.}
    \setlength{\belowcaptionskip}{-10pt}
    \caption{\textcolor{black}{Sensitivity analysis of the HITL framework under different maximum probability caps $p^\text{max}$ with discomfort proportion $\beta$ = 0.5.}}
    \label{fig:sensitivity_analysis}
\end{figure}

\textcolor{black}{Figure \ref{fig:sensitivity_analysis} illustrates our framework's sensitivity to imprecise human feedback across different maximum probability caps ($p^\text{max}$ from 0.50 to 1), where we present results with discomfort proportion $\beta = 0.5$ to demonstrate the system's performance when balancing comfort and energy efficiency. The total cost across scenarios remains relatively stable, though showing slight decreases at lower probability caps due to reduced feedback frequency leading to lower discomfort costs. Temperature control performance, measured by MAE to setpoint, demonstrates robust stability across all scenarios. Even in S2 (no occupancy information), which shows the highest sensitivity, the maximum MAE fluctuation is around 12\% across the probability range due to its complete reliance on feedback without occupancy inference. S3 and S4 maintain relatively stable performance as they benefit from either current or predicted occupancy information to complement the feedback mechanism. S1, with perfect occupancy prediction, exhibits the most stable performance with variations of less than 0.5\% in MAE, demonstrating that accurate occupancy and prediction information can effectively compensate for imprecise feedback patterns.}

\section{Conclusion}
This paper presents a Human-in-the-Loop (HITL) AI framework that simplifies the complexity of existing methods for optimizing HVAC systems while effectively balancing energy efficiency and occupant comfort. By incorporating real-time user feedback and responding to dynamic electricity market conditions, the proposed framework demonstrates significant improvements in reducing energy costs and enhancing comfort without relying on the complex predefined models that characterize much of the existing work. The use of RL allows the system to continuously adapt to changing environments and occupant preferences, offering a scalable and flexible solution. 
\textcolor{black}{In future work, we will explore further refinement of occupancy prediction and scalability across various building types and grid conditions, with particular attention to privacy-preserving techniques for occupancy data collection and processing. Additionally, future research could investigate the implementation of delayed feedback mechanisms within timesteps to more accurately model real-world user interaction patterns. While this study demonstrates the effectiveness of occupancy-aware HVAC control, implementing such systems in practice will require careful consideration of user privacy alongside energy efficiency and comfort optimization.}

\begin{acks}
This work is supported in parts by the Australian Research Council (ARC) through Discovery Early Career Researcher Award (DECRA) under Grant DE230100046 and the ARC Training Centre in Optimisation Technologies, Integrated Methodologies and Applications (OPTIMA) under Grant IC200100009.
\end{acks}

\newpage

\bibliographystyle{ACM-Reference-Format}
\bibliography{reference}

\end{document}